\documentclass[preprint]{aastex}
\received{2005 January 3}
\begin{document}

\title{Star Formation Histories of Nearby Elliptical Galaxies. II. 
Merger Remnant Sample}
\author{Justin H.\ Howell}
\affil{UCO/Lick Observatory, Department of Astronomy \& Astrophysics,\\
University of California, Santa Cruz, California 95064, USA\\
\tt{jhhowell@ucolick.org}\footnote{now at the Infrared Processing and
Analysis Center, Mail Stop 100-22, California Institute of Technology, Jet
Propulsion Laboratory, Pasadena, CA 91125; jhhowell@ipac.caltech.edu}}


\begin{abstract}

This work presents high $S/N$ spectroscopic observations of a
sample of six suspected merger remnants, selected primarily on the basis of 
H{\sc i} tidal debris detections.  Single stellar population analysis of these
galaxies indicates that their ages, metallicities, and $\alpha$-enhancement
ratios are consistent with those of a representative sample of nearby
elliptical galaxies.  The expected stellar population of a recent merger
remnant, young age combined with low [$\alpha$/Fe], is not seen in any
H{\sc i}-selected galaxy.  However, one galaxy (NGC~2534), is found to 
deviate from the $Z$-plane in the sense expected for a merger remnant. 
Another galaxy (NGC~7332), selected by other criteria, best matches the
merger remnant expectations.

\end{abstract}


\section{Introduction}

The formation mechanisms for early-type galaxies have been the subject
of intense scrutiny over the past several decades.  The early dynamical
simulations of \citet{ch2toomre72} showed that merger events between two
spiral galaxies can result in an elliptical galaxy as the merger remnant.
This provides a striking alternative to the classical monolithic collapse
picture of elliptical galaxy formation \citep[e.~g.][]{larson}, in which
the entire galaxy forms in a single star-formation event at high redshift.
Although mergers in progress are obvious, merger remnants evolving into 
elliptical galaxies are more difficult to identify.  Such objects populate
the ``King gap'' \citep{king} separating interacting galaxies from the
quiescent elliptical remnant galaxy they are proposed to become.
For purposes of this study, major mergers are defined as those
encounters in which the mass ratio is large enough that any stellar disks
are disrupted and transformed into an elliptical merger remnant.  Minor
mergers or accretion events are defined as encounters in which the structure
of the higher-mass progenitor (whether elliptical or spiral) is preserved 
in the remnant.

The predicted stellar populations of merger remnants depend greatly on the
gas fraction in the progenitor galaxies.  In the extreme case suggested
by \citet{ashzepf}, where ellipticals form from predominantly gaseous spiral
galaxies, the resulting stellar population will differ from that of a 
monolithic collapse elliptical only in age.

An alternative expectation for the stellar population of the remnant
of a recent major merger event is similar to the frosting model of
\citet{ch2tfwg2}.  The starburst at the time of the merger event involves
only a small fraction of the total mass of the system, with the bulk of 
the galaxy being older and less metal-rich.  A key difference between the 
expectations of such a merger remnant and a an elliptical formed in a
single starburst is that the latter will have high [$\alpha$/Fe] due to 
the short formation timescale, while stars formed within two spiral 
galaxies will have near solar [$\alpha$/Fe] \citep*{tgb99}.  In a merger
between spiral galaxies which have already converted much of their mass
into stars, the burst population will also have low 
$[\alpha/{\rm Fe}]\leq0.1$, as the gas from which it forms 
has been enriched with iron from billions of years of SN~Ia.  \citet{tgb99} 
showed that although it is possible for a merger scenario to produce the 
high [$\alpha$/Fe] values seen in most early-type galaxies, a flat initial 
mass function (IMF) for the merger starburst population is required in 
order to produce sufficient $\alpha$-elements.  This 
prediction of low [$\alpha$/Fe] is generalized to any galaxy formed 
hierarchically by \citet{thomas99} using the semianalytic models of 
\citet{k96}.  The recent work of \citet{nagashima} arrives at the same 
conclusions, obtaining enhanced [$\alpha$/Fe] only for a top-heavy IMF, 
or near solar [$\alpha$/Fe] using an IMF more typical of quiescent star 
formation in spiral galaxies.

The paper is organized as follows.  The sample selection and data are
described in \S~2.  Stellar population analysis is performed in \S~3,
and conclusions are discussed in \S~4.

\section{Sample Selection and Data}

The bulk of the merger remnant sample was selected based on detections of 
H{\sc i} tidal debris (J.~Hibbard 2000, private communication).  As 
\citet{hibmihos} showed, gas in tidal tails can remain bound but at large 
radii for several Gyr after a major merger event.  This criterion 
preferentially selects early-type galaxies resulting from gas-rich
encounters.  Galaxies which accreted a gas-rich satellite are
the primary contaminants for a sample of major merger remnants using this 
H{\sc i} selection criterion.  For example, \citet{vg86} proposed such an 
accretion event to explain the H{\sc i} observations of NGC~1052, and
\citet{vdriel89} made a similar argument about NGC~3619.
Other selection criteria, such as the fine structure parameter of
\citet{ss92}, are sensitive to all types of merger event, major or minor,
gas-rich or gas-poor.  \citet{hibsans} have shown that galaxies with large
fine structure parameters are no more likely than those with little fine
structure to have associated H{\sc i} tidal debris.
One galaxy in the selected sample, NGC~1052, is also
a member of the volume-limited sample of \citet[][hereafter Paper I]{paper1}
The merger remnant sample is listed in Table~1.  Morphological T-types 
\citep{rc3} and the $S/N$ ratio per pixel
near the H$\beta$ line are listed for each galaxy.

NGC~7332 does not satisfy the H{\sc i} selection criterion \citep{hi7332} but
shows other indications suggesting a merger origin.  Deviations from
the Fundamental Plane \citep{prugniel, forbes98} suggest a young age, and
\citet{ss92} find substantial fine structure ($\Sigma=4.00$) and a young
heuristic merger age $\sim5$~Gyr.  Although NGC~7332 has a bimodal color
distribution of globular clusters, the age of the red population has not
been spectroscopically determined.  The galaxy itself has been studied
spectroscopically; see \citet{tf02} for a summary.  The SAURON group has
recently observed the entire galaxy at a limited wavelength coverage
\citep{sauron04}.  Both spectroscopic studies find consistent ages of
4.5--5~Gyr.

Optical images overlaid with H{\sc i} maps are presented for five galaxies
in the merger remnant sample in Figs.~\ref{474map}--\ref{5903map}.  NGC~474
is a gas-poor shell galaxy interacting with the nearby gas-rich spiral 
galaxy NGC~470.  It has a very large fine structure index $\Sigma=5.26$
and a relatively young heuristic merger age $\sim4$~Gyr \citep{ss92}.
However, the heuristic merger age is based on an analysis of $UBV$
color information, which is known to be insufficient to break the 
age-metallicity degeneracy \citep{w94model}.  NGC~1052 is a prototypical LINER
system \citep{1052liner} with two H{\sc i} tails (Fig.~\ref{1052map}).
\citet{ss92} find little fine structure ($\Sigma=1.78$) and a heuristic
merger age $\sim8$~Gyr.  NGC~2534 has a single extended H{\sc i} tail, 
as shown in Fig.~\ref{2534map}.  In NGC~3619, the H{\sc i} emission is
colocated with the galaxy's stellar component, with a smaller 
characteristic radius (Fig.~\ref{3619map}).  
NGC~5903 shows extended H{\sc i} both within the galaxy and in two long
tidal tails (Fig.~\ref{5903map}).

By construction, this sample is expected to have young ages as noted 
above.  Assuming that these galaxies formed in major merger events (and 
that the merger starburst did not have an IMF strongly biased towards 
massive stars), they are also expected to have low [$\alpha$/Fe] \citep{tgb99}.

Observations and data analysis were performed identically as for the 
volume-limited sample described in Paper~I.  Longslit spectra were taken
using the Kast spectrograph with the 1200 lines/mm grating blazed at 
$5000\mbox{\AA}$ on the Lick 3~meter telescope.  The $145''$-long slit 
was oriented along the galaxy's major axis.  Four 25~minute exposures
were taken on each galaxy, interspersed with 5~minute sky exposures taken
several arcminutes away.  The spectral range was 
$4200\mbox{\AA}$--$5600\mbox{\AA}$, with instrumental resolution of 
approximately 100~km/s.  The slit width was typically $1.5^{\prime\prime}$,
though in poor seeing conditions this was increased to $2^{\prime\prime}$.
The plate scale for the spectrograph in this configuration was
$1.17\mbox{\AA}$/pixel in the dispersion direction and
$0.8^{\prime\prime}$/pixel in the spatial direction.

The data were flat fielded, masked for cosmic rays and bad pixels, and
combined into a single spectrum for each galaxy.  For most galaxies, sky 
subtraction was performed using the edges of the slit, though some used
the sky exposures instead as described in Paper~I.  None of the galaxies
in this sample are large enough in angular size for light from the galaxy
itself to significantly bias the sky measurement at the slit edges.  
The galaxy spectra were extracted in $r_e/8$ apertures, with effective 
radii taken from \citet{faber89} where possible, and from \citet{rc3} 
otherwise (NGC~2534).  Equivalent widths for the available Lick indices were 
calculated using a version of the {\tt bwid} program, provided by 
R.~M.~Rich \citep{bwid}.  Velocity dispersion and emission corrections 
were performed using the standard methods described in Paper~I.  Emission 
corrections were non-negligible in all galaxies in this sample, as 
expected for objects with young stellar populations.  Finally, the 
standard star observations detailed in Paper~I were used to transform the 
index measurements onto the Lick/IDS system.  Index errors were calculated 
using the error simulations and systematic uncertainties derived in Paper~I.
Measurements for all available indices are presented in Table~2.

\section{Analysis}

\subsection{Single Stellar Populations}

One set of stellar population parameters was derived from the primary indices, 
H$\beta$, Mg$b$, Fe5270, and Fe5335.  Models from \citet*{ch2tmb03} were
used to interpolate age, [Z/H], and [$\alpha$/Fe] in the 
H$\beta$--[MgFe]$^\prime$ (Fig.~\ref{hbmr}) and Mg$b$--$\langle{\rm Fe}\rangle$ 
planes.  The H$\gamma$--[MgFe]$^\prime$ plane can also be 
used to interpolate stellar population parameters, using models designed for 
the higher-order Balmer lines by \citet*{tmk04} (Fig.~\ref{hgmr}).
A discussion of the relative merits of the two Balmer lines as
age indicators is presented in Paper~I.  Since the merger remnant sample
suffers from considerable emission contamination, the ages derived using
H$\gamma$ as the age-sensitive index are expected to be more
accurate than ages using H$\beta$ as the age-sensitive index.  
Quoted uncertainties are derived from the 
index uncertainties in the manner described in Paper~I; note that the
resulting SSP uncertainties do not incorporate the systematic uncertainties
associated with the choice of specific models and indices used to 
measure SSP quantities.  

Ideally, SSP parameters should be measured using all available information, 
not just the handful of commonly-used indices described above.  The 
multiple index fitting code of \citet[][hereafter P04]{proctor04} provides 
the most reliable measurements of age, [Z/H], and $[\alpha/{\rm Fe}]$, fitting 
the three SSP parameters simultaneously using every index measured from 
the galaxy spectrum.  Of particular importance for the present 
sample, deviant indices such as Balmer lines with large emission corrections 
can be omitted from the fit.  The drawback of this method is that it 
relies on a different calibration of $[\alpha/{\rm Fe}]$ than the 
Thomas~et~al. models.  Both Thomas~et~al. and P04 take 
into account the variation of $[\alpha/{\rm Fe}]$ with [Fe/H] below 
solar metallicity in the stellar calibrators used in the construction of SSPs.
P04 extends the correction to supersolar metallicities.
This seemingly minor difference has major effects on correlations 
between SSP parameters.  P04 showed that using the 
corrected calibration, the well-known $[\alpha/{\rm Fe}]$--$\sigma$
relation is destroyed, while correlations between $[\alpha/{\rm Fe}]$ 
and age, and $[\alpha/{\rm H}]$ and $\sigma$ are strengthened.  An 
unfortunate consequence of the calibration difference between 
P04 and other models is that the results of the 
multiple index fitting code cannot be directly compared to previous 
results.

The SSP parameters measured using all three 
methods are listed in Table~3; both the SSP values derived using H$\gamma$ 
as the age-sensitive index and multi-index fitting values will be used 
in the subsequent analysis.  Good fits were obtained for all galaxies 
using the latter method (Proctor 2005, private communication), though the 
fits for NGC~474 and NGC~1052 were noticeably worse than the rest.

Stellar population parameters age, [Z/H], and [$\alpha$/Fe] are 
plotted against structural parameters $\sigma$, $M_B$, and log~$r_e$
in Fig.~\ref{asigmr}.  
Also shown are the galaxies from the 
volume limited sample described in Paper~I and galaxies from the 
\citet[][hereafter G93]{ch2g93}
sample not included in the volume-limited sample.  Index measurements
from G93 have been used to measure SSP parameters using the \citet{ch2tmb03} 
models for consistency.  The stellar population measurements from the 
latter samples are derived from H$\beta$--[MgFe]$^\prime$ model grids.
However, for the purpose of illustrating the underlying qualitative trends 
within a large and representative population of elliptical galaxies, the 
difference between SSP measurements using H$\gamma$ or H$\beta$ as the
age-sensitive index is not of critical importance.  Due to the difference 
in [$\alpha$/Fe] calibration, the results of the multi-index fitting are
not as directly comparable.
Figure~\ref{asigmr} shows that the galaxies in the merger remnant sample
have the same distribution of stellar population parameters as
galaxies from the volume-limited sample with similar structural parameters.
According to Kolmogorov-Smirnov tests, the probabilities that the samples 
are drawn from the same distribution are 0.997, 0.59, and 0.24 in age, 
[Z/H], and [$\alpha$/Fe] respectively.  
With four of six galaxies (all six, using the multi-index fitting method) 
having SSP ages less than or equal to 6~Gyr,
the merger remnant sample is, on average, significantly younger than the 
volume-limited sample as a whole.  The young ages cannot be directly 
attributed to mergers, however, since these galaxies are on the low 
mass, faint, small radius end of the distribution of the volume-limited 
sample.  The similarly young ages of galaxies in the volume-limited sample 
with comparable sizes and luminosities suggests downsizing \citep{cowie}
as an adequate explanation.

The merger remnant sample is also plotted on the metallicity hyperplane
in Fig.~\ref{pcmr}.  This hyperplane was derived by \citep{ch2tfwg2} using 
principal component analysis in the four dimensional parameter space of 
velocity dispersion, age, metallicity, and [$\alpha$/Fe] abundance ratio.
PC1 increases with $\sigma$, 
[$\alpha$/Fe], and to a lesser extent [Z/H]; PC2 increases with age and
decreases with [Z/H]; and PC3 increases with [$\alpha$/Fe] and decreases
with $\sigma$.  Thus the remnants of major mergers, having low ages and
[$\alpha$/Fe], would be expected to have low values in each principal
component.   No galaxy in the merger remnant sample occupies that part of
the hyperplane.  The unusually low PC1 value of NGC~7332 is a result of a
small velocity dispersion combined with low [$\alpha$/Fe].  The remaining 
galaxies in the sample are distributed similarly to the galaxies in the
volume-limited sample.  Note that the calibration difference between 
the multi-index fitting technique and previous SSP model estimates 
prevents any meaningful use of the multi-index fits in reference to 
the hyperplane.  As mentioned previously, the [$\alpha$/Fe]--$\sigma$ 
relation (represented by the first principal component in the \citet{ch2tfwg2}
hyperplane) disappears using the P04 calibration.

The distribution of the merger remnant sample along the $Z$-plane 
\citep{ch2tfwg2} is shown 
in Fig.~\ref{zplanemr}.  Note that as described in Paper~I the best fit 
line must be offset slightly in [Z/H] to account for the use of different 
models than in \citet{ch2tfwg2}.  \citet{ch2tfwg2} argued that a scaling 
between age and metallicity can most sensibly be maintained over time by 
episodes of star formation within the host galaxy.  Stellar populations
formed in a single starburst cannot maintain a linear $Z$-plane projection
in this space of [Z/H], log~$t$, and log~$\sigma$ since log~$t$ changes more
rapidly at younger ages than at older ages.  A major merger between 
two spiral galaxies should not obey this same relation barring an improbable 
conspiracy of stellar population parameters in the progenitor galaxies.
The metallicity and abundance ratios of the stars formed
in the merger starburst must have the proper scaling with respect to the
existing stellar populations of the progenitor galaxies in order for the
merger remnant to return to the $Z$-plane and the metallicity hyperplane.
Further, the range of allowed metallicities for the starburst population
is smaller for younger remnants than for older remnants.  As with the 
hyperplane above, multi-index fits cannot be meaningfully plotted on 
the $Z$-plane.

NGC~2534 is the only outlier from the $Z$-plane defined by the galaxies in 
the volume-limited sample.  The measured [Z/H] for this galaxy is too low 
by $\sim0.3$~dex for it to lie along the plane; alternatively an age 
older by $\sim3.5$~Gyr would bring NGC~2534 onto the $Z$-plane 
at the present [Z/H] value.  The latter is the more physically plausible 
explanation: a small mass fraction starburst can easily decrease the SSP 
age measurement by that amount and shift a galaxy off of the $Z$-plane 
for several Gyr.  Whether or not passive evolution ages the galaxy back 
to the $Z$-plane depends on the metallicities of the original and starburst 
populations \citep{ch2tfwg2}.  A major merger event 
could explain the location of NGC~2534 in this parameter space; this 
possibility will be discussed in detail in \S~4.  It is interesting that 
the other five galaxies all lie on the $Z$-plane in good agreement 
with the galaxies from Paper~I and G93.  Since the multi-index fits are 
not usable for this analysis, the errors for NGC~1052 are extremely large 
and it is not a significant outlier.
As candidate merger remnants, 
these galaxies would also be expected to lie above this projection of the 
plane for the
same reasons as NGC~2534.  It is worth noting that although galaxies which
lie above the $Z$-plane are likely to have a frosting population of
young stars, not all galaxies with such a frosting population will necessarily
lie above the $Z$-plane.  However, gas originating in two distinct galaxies 
is unlikely to have the necessary scaling of metal abundances to return the 
resulting merger remnant to the $Z$-plane.  The fact that so many suspected
merger remnants --- NGC~474, NGC~3619, NGC~5903, NGC~7332; NGC~3610 
\citep{3610}; NGC~584, NGC~1700, NGC~5831, NGC~6702 \citep{ch2tfwg2} --- 
do lie along the $Z$-plane is therefore an intriguing mystery.
Alternatively, this may indicate that the $Z$-plane is not as useful a 
discriminant between past star formation histories as it appears, since
almost all ellipticals regardless of formation mechanism lie on this plane.

\section{Discussion and Conclusions}

A small sample of merger remnant galaxies has been selected based primarily on
evidence of H{\sc i} tidal debris.  High quality spectral line index 
measurements have been used to estimate stellar population parameters (age, 
metallicity, $\alpha$-enhancement) using the best available SSP models 
and multi-index fitting techniques.

Four galaxies, NGC~474, NGC~1052, NGC~3619, and NGC~5903, have broadly
similar properties.  All lie along the $Z$-plane as defined by the 
volume-limited sample of Paper~I.  All have intermediate SSP ages 
(4--6~Gyr) as measured using the multi-index fitting method of P04.
These galaxies have stellar populations more consistent with the 
volume-limited sample (Paper~I) than with predictions of recent
major merger remnants.  The stellar population measured for NGC~1052 can be 
compared with the definitive study of \citet{pierce}.  The results of 
the multi-index fit are in reasonable agreement with \citet{pierce}, 
despite the fact that measurements relying on either Balmer line are 
extremely inaccurate due to the strong emission contamination.

The young SSP age and relatively low [Z/H] of NGC~2534 make it an outlier 
from the $Z$-plane in precisely the sense
one would expect of a merger remnant with a small (by mass fraction) young
population overlying an older population.  However, the [$\alpha$/Fe]
ratio in NGC~2534 is very large for an object in which most of the stars
formed in spiral galaxies, or from gas enriched with iron by the long 
continuous star formation typical of spirals.  Taking into account the
relatively small velocity dispersion, NGC~2534 is somewhat {\it more}
$\alpha$-enhanced than most elliptical galaxies (Fig.~\ref{asigmr}).  
This high [$\alpha$/Fe] ratio places the galaxy at a very high PC3 value 
in the metallicity hyperplane.
Instead of the remnant of a major merger, these results are more consistent
with NGC~2534 being an old, pre-existing elliptical which recently accreted
a small, gas-rich companion.

NGC~7332 is another good match to merger remnant expectations.  The SSP 
measurements for NGC~7332 are in excellent agreement with other studies
\citep{sauron04, tf02} when the H$\gamma$ model grids are used.  The 
multi-index fit presented here yields a significantly younger age and higher 
metallicity.  The extreme position of NGC~7332 in Fig.~\ref{pcmr} is due to 
the combination of low velocity dispersion and low (near solar) [$\alpha$/Fe].
The measurement of $[\alpha/{\rm Fe}] = +0.15$ (multi-index fit) is somewhat 
larger than predicted \citep{tgb99}, and is also consistent with galaxies 
of similar size in the volume-limited sample.
As discussed above and in \citet{ch2tfwg2},
the metallicities of the progenitor galaxies would have to be carefully 
matched for the merger remnant to lie along the $Z$-plane as NGC~7332 does.

Fundamental Plane residuals provide an independent indication of anomalously 
young ages such as would be produced by a merger-induced starburst 
\citep{forbes98}.  The study of \citet{rothberg05} showed that optically 
selected merger remnants which deviate from the Fundamental Plane do so 
only in having higher surface brightness.
\citet{prugniel} calculated Fundamental Plane residuals for five 
galaxies in the present sample.  The residual for NGC~2534 was calculated
using Equation~4 of \citet{prugniel} with data taken from \citet{rc3}. 
The Fundamental Plane residuals of NGC~474, 1052, 2534, 3619, and 5903 
range from $+0.03$~to~$+0.11$.  Only NGC~7332 has a large negative 
residual indicative of a young age \citep{forbes98}.

To summarize, two of the six galaxies fit some of the expected properties 
of a ``King gap'' merger remnant \citep{tgb99}, though neither galaxy 
precisely fits every prediction.  The positions of NGC~7332 on the 
hyper-plane and relative to the Fundamental Plane suggest that it is 
the most likely merger remnant in the sample, though the abundance 
ratio [$\alpha$/Fe] (multi-index fit) and position on the $Z$-plane 
remain difficult to 
explain.  Taken as a group the SSP properties of the merger remnant 
sample are not substantially different from those of structurally
similar (in velocity dispersion, luminosity, or radius) galaxies from the
volume-limited sample of Paper~I (Fig.~\ref{asigmr}).  The merger remnant
sample is consistent with being drawn from the same distribution as the
volume-limited sample in each SSP parameter. 

The disagreement between derived SSP quantities and merger remnant predictions
should not necessarily be taken to imply that the galaxies are not merger 
remnants.  The similarities between the properties
of the merger remnant sample and the volume-limited sample, as well as the
non-detection of the expected major merger signatures (low age combined with
near solar [$\alpha$/Fe]) in four of the six galaxies suggest that this 
``merger remnant'' sample may include few remnants of recent major mergers.  
Instead, as proposed for NGC~2534, these galaxies are more consistent with 
being older elliptical galaxies which recently accreted a gas-rich companion 
galaxy.  It is also possible that the predictions for the stellar populations 
of major merger remnants are not applicable to the galaxies in this sample. 
\citet{tgb99} consider a merger of two galaxies similar to the Milky Way; 
more gas-rich progenitors and thus a larger starburst by mass fraction would 
result in supersolar [$\alpha$/Fe].  A top-heavy IMF has long been suggested 
as another explanation for the large observed values of [$\alpha$/Fe] in 
early-type galaxies.  \citet{nagashima} incorporated such an IMF into 
semi-analytical hierarchical galaxy formation models which reproduce the 
observed [$\alpha$/Fe] values for some elliptical galaxies, though the 
models at present fail to reproduce the [$\alpha$/Fe]-$\sigma$ relation.
As P04 discussed extensively, the calibration of 
[$\alpha$/Fe] is tremendously important since many galaxy formation 
models use the [$\alpha$/Fe]--$\sigma$ relation as a constraint.

\acknowledgments

This research has made use of the NASA/IPAC Extragalactic Database (NED) 
which is operated by the Jet Propulsion Laboratory, California Institute of 
Technology, under contract with the National Aeronautics and Space 
Administration.  This work also made use of the Gauss-Hermite Pixel
Fitting Software developed by R.P. van der Marel.  J.~H.~H. was supported 
in part by an ARCS Fellowship.  This research was supported in part by 
NSF grant AST-0507483 to the University of California Santa Cruz.  We 
thank Ricardo Schiavon, Sandy Faber,
Jean Brodie, and Mike Beasley for helpful conversations, Rob Proctor 
for the use of his multi-index fitting code and several important comments, 
and the anonymous referee for comments which significantly improved the paper.

\vfill\eject

\begin{deluxetable}{lllllll}
\tablecolumns{7}
\tablewidth{0pt}
\tablecaption{Merger Remnant Galaxy Sample} 
\tablehead{
\colhead{Name} & \colhead{$\alpha$(J2000)} & \colhead{$\delta$(J2000)} & \colhead{T-type} & \colhead{$r_e$}& \colhead{$S/N$} & \colhead{Observed} \\
}
\startdata
NGC 474 & $01^{\rm h}20^{\rm m}06.70^{\rm s}$ &  $+03^{\circ}24^{\prime}55.0^{\prime\prime}$ & $-2$ & $33.7^{\prime\prime}$ & 133.9 & 2000 November 27 \\
NGC 1052 &  $02^{\rm h}41^{\rm m}04.80^{\rm s}$ &  $-08^{\circ}15^{\prime}20.8^{\prime\prime}$ & $-5$ & $36.9^{\prime\prime}$ & 205.8 & 2000 November 26 \\
NGC 2534 &  $08^{\rm h}12^{\rm m}54.1^{\rm s}$ &  $+55^{\circ}40^{\prime}19^{\prime\prime}$ & $-5$ &  $20.3^{\prime\prime}$ & 66.0 & 2002 November 3\\
NGC 3619 & $11^{\rm h}19^{\rm m}21.6^{\rm s}$ &  $+57^{\circ}45^{\prime}29^{\prime\prime}$ & $-1$ & $25.5^{\prime\prime}$ & 107.3 & 2001 March 26 \\
NGC 5903 & $15^{\rm h}18^{\rm m}36.3^{\rm s}$ &  $-24^{\circ}04^{\prime}06^{\prime\prime}$ & $-5$ & $35.3^{\prime\prime}$ & 88.9 & 2001 March 25 \\
NGC 7332 & $22^{\rm h}37^{\rm m}24.5^{\rm s}$ &  $+23^{\circ}47^{\prime}54^{\prime\prime}$ & $-2$ & $14.7^{\prime\prime}$ & 146.8 & 2002 November 2\\
\enddata
\end{deluxetable}

\begin{deluxetable}{lrllrlllllrllllll}
\tabletypesize{\scriptsize}
\rotate
\tablecolumns{17}
\tablewidth{0pt}
\tablecaption{Index Measurements: $r_e/8$ Aperture} 
\tablehead{
\colhead{Galaxy} & \colhead{$\sigma$} & \colhead{Ca4227} & \colhead{G4300} & \colhead{H$\gamma_{\rm F}$} & \colhead{Fe4383} & \colhead{Ca4455} & \colhead{Fe4531} & \colhead{C4668} & \colhead{H$\beta$} & \colhead{[OIII]$\lambda5007$} & \colhead{Fe5015} & \colhead{Mg$b$} & \colhead{Fe5270} & \colhead{Fe5335} & \colhead{Fe5406} & \colhead{[MgFe]$^\prime$} \\
}
\startdata
NGC 474 & 169 & 1.12 & 4.75 & -1.05 & 5.34 & 1.58 & 3.75 & 6.79 & 1.86 & -0.49~~~~~~ & 6.12 & 4.55 & 3.25 & 2.55 & 1.56 & 3.73 \\
~~~ & 4 & 0.10 & 0.14 & 0.11 & 0.09 & 0.05 & 0.11 & 0.12 & 0.07 & 0.07~~~~~~ & 0.20 & 0.05 & 0.06 & 0.09 & 0.09 & 0.05 \\
NGC 1052 & 215 & 1.20 & 5.72 & -2.17 & 6.50 & 1.43 & 3.78 & 8.24 & 1.21 & -3.71~~~~~~ & 1.90 & 5.96 & 3.05 & 2.78 & 1.88 & 4.21 \\
~~~ & 4 & 0.10 & 0.14 & 0.38 & 0.09 & 0.05 & 0.11 & 0.12 & 0.7 & 1.0~~~~~~ & 0.64 & 0.05 & 0.06 & 0.09 & 0.09 & 0.05 \\
NGC 2534 & 158 & 0.91 & 4.25 & 0.01 & 3.90 & 1.17 & 3.24 & 6.12 & 2.27 & -1.02~~~~~~ & 4.74 & 3.70 & 2.56 & 2.12 & 1.50 & 3.00 \\
~~~ & 6 & 0.10 & 0.14 & 0.13 & 0.21 & 0.14 & 0.09 & 0.20 & 0.11 & 0.14~~~~~~ & 0.29 & 0.06 & 0.09 & 0.13 & 0.09 & 0.07 \\
NGC 3619 & 179 & 1.08 & 5.56 & -1.33 & 5.81 & 1.43 & 3.88 & 7.48 & 2.25 & -1.25~~~~~~ & 4.63 & 4.47 & 3.19 & 2.99 & 1.96 & 3.74 \\
~~~ & 4 & 0.10 & 0.16 & 0.10 & 0.18 & 0.07 & 0.07 & 0.19 & 0.12 & 0.16~~~~~~ & 0.14 & 0.06 & 0.09 & 0.13 & 0.09 & 0.07 \\
NGC 5903 & 211 & 1.40 & 5.97 & -2.10 & 5.73 & 1.73 & 3.72 & 7.00 & 1.92 & -0.42~~~~~~ & 5.83 & 4.57 & 3.05 & 2.78 & 1.63 & 3.69 \\
~~~ & 4 & 0.10 & 0.16 & 0.08 & 0.18 & 0.07 & 0.07 & 0.19 & 0.08 & 0.08~~~~~~ & 0.12 & 0.06 & 0.09 & 0.13 & 0.09 & 0.07 \\
NGC 7332 & 128 & 1.14 & 5.14 & -1.07 & 4.66 & 1.56 & 3.74 & 7.43 & 2.47 & -0.32~~~~~~ & 6.09 & 3.83 & 3.13 & 2.84 & 1.87 & 3.42 \\
~~~ & 4 & 0.10 & 0.14 & 0.13 & 0.21 & 0.14 & 0.09 & 0.20 & 0.09 & 0.10~~~~~~ & 0.29 & 0.05 & 0.06 & 0.09 & 0.09 & 0.05 \\
\enddata
\end{deluxetable}
\normalsize

\begin{deluxetable}{llllllllll}
\tablecolumns{10}
\tablewidth{0pt}
\tablecaption{Stellar Population Parameters} 
\tablehead{
\colhead{Galaxy} & \colhead{$\sigma$} & \colhead{$\sigma_{\sigma}$} & Method & \colhead{Age} & \colhead{$\sigma_t$} & \colhead{[Z/H]} & \colhead{$\sigma_Z$} & \colhead{$[\alpha/{\rm Fe}]$} & \colhead{$\sigma_{\alpha}$} \\
        & km/s & & & Gyr & & dex & & dex \\
}
\startdata
NGC 474 & 169 & 4 & H$\beta$ grid & 3.4 & 0.6 & 0.55 & 0.09 & 0.27 & 0.04 \\
& & & H$\gamma$ grid & 3.45 & 0.8 & 0.53 & 0.07 & 0.27 & 0.04 \\
& & & Multi-index fit & 4.4 & 1.2 & 0.38 & 0.07 & 0.24 & 0.03 \\
NGC 1052 & 215 & 5 & H$\beta$ grid & 16 & 14 & 0.42 & 0.3 & 0.44 & 0.12 \\
& & & H$\gamma$ grid & 10.9 & 8 & 0.48 & 0.16 & 0.46 & 0.09 \\
& & & Multi-index fit & 4 & 2 & 0.45 & 0.08 & 0.21 & 0.05 \\
NGC 2534 & 158 & 6 & H$\beta$ grid & 2.2 & 0.9 & 0.23 & 0.08 & 0.33 & 0.07 \\
& & & H$\gamma$ grid & 2.8 & 0.7 & 0.17 & 0.06 & 0.30 & 0.07 \\
& & & Multi-index fit & 2.7 & 0.2 & 0.20 & 0.05 & 0.27 & 0.04 \\
NGC 3619 & 179 & 4 & H$\beta$ grid & 1.2 & 1 & 0.94 & 0.16 & 0.33 & 0.05 \\
& & & H$\gamma$ grid & 5.4 & 1.5 & 0.43 & 0.06 & 0.18 & 0.05 \\
& & & Multi-index fit & 4.0 & 0.8 & 0.43 & 0.06 & 0.15 & 0.03 \\
NGC 5903 & 211 & 4 & H$\beta$ grid & 3.0 & 0.5 & 0.58 & 0.09 & 0.28 & 0.04 \\
& & & H$\gamma$ grid & 17.5 & 2 & 0.16 & 0.1 & 0.18 & 0.04 \\
& & & Multi-index fit & 6.0 & 1.6 & 0.28 & 0.04 & 0.15 & 0.05 \\
NGC 7332 & 128 & 4 & H$\beta$ grid & 0.8 & 0.3 & 0.88 & 0.14 & 0.26 & 0.04 \\
& & & H$\gamma$ grid & 6.1 & 1 & 0.22 & 0.06 & 0.06 & 0.04 \\
& & & Multi-index fit & 1.8 & 0.2 & 0.58 & 0.09 & 0.15 & 0.03 \\
\enddata
\end{deluxetable}

\vbox{
\begin{center}
\leavevmode
\includegraphics[width=\textwidth]{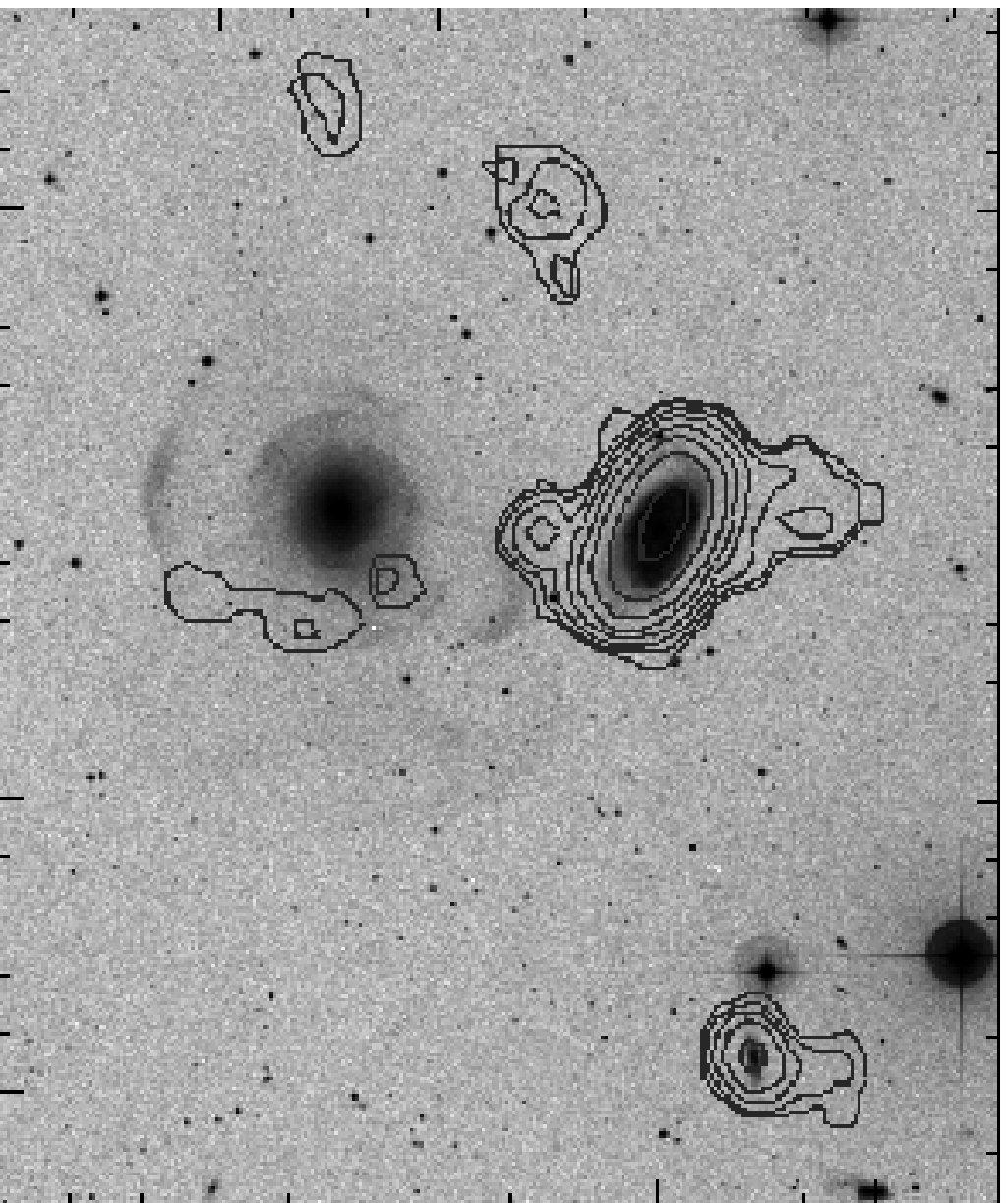}
\figcaption{\small
Contours of H{\sc i} emission are overlaid on the DSS image of 
NGC~474 (left) and the nearby spiral galaxy NGC~470 (right).  The
disturbed, shell galaxy morphology of NGC~474 is clearly evident.
Although NGC~474 has little or no H{\sc i} of its own, it appears 
to have accreted gas from NGC~470.  This figure originally appeared
in \citet{dave}.
\label{474map}
}
\end{center}}

\vbox{
\begin{center}
\includegraphics[width=\textwidth]{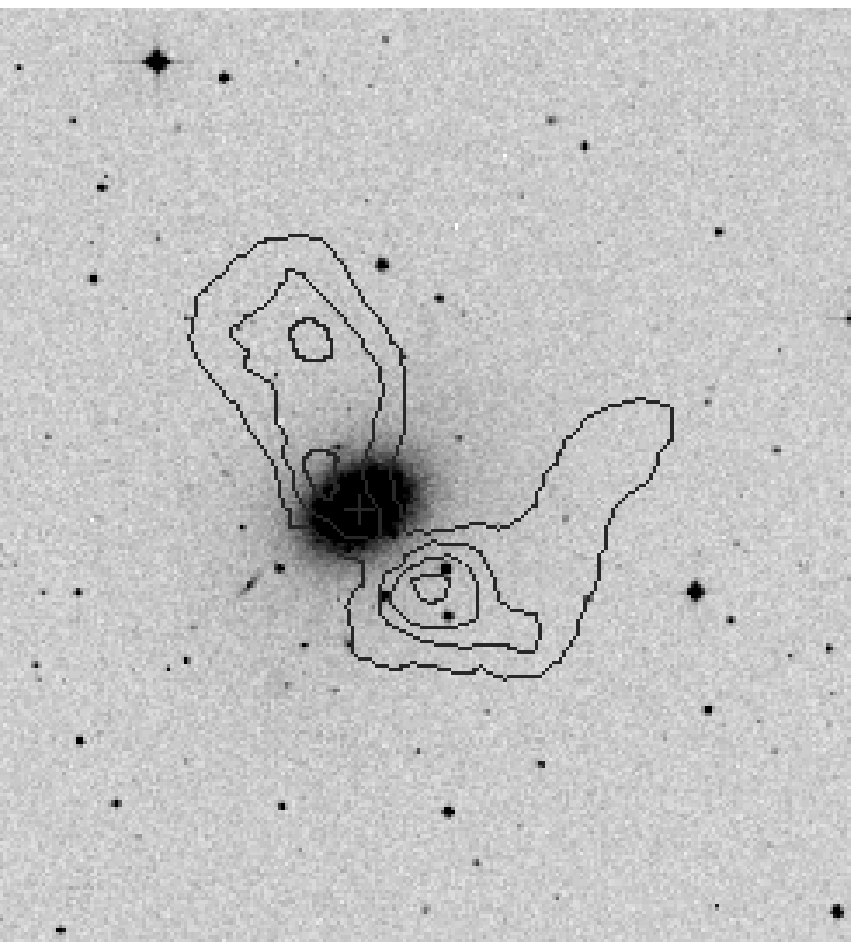}
\figcaption{\small
Contours of H{\sc i} emission are overlaid on the DSS image of 
NGC~1052.  This figure originally appeared in \citet{vg86}.
\label{1052map}
}
\end{center}}

\vbox{
\begin{center}
\includegraphics[width=\textwidth]{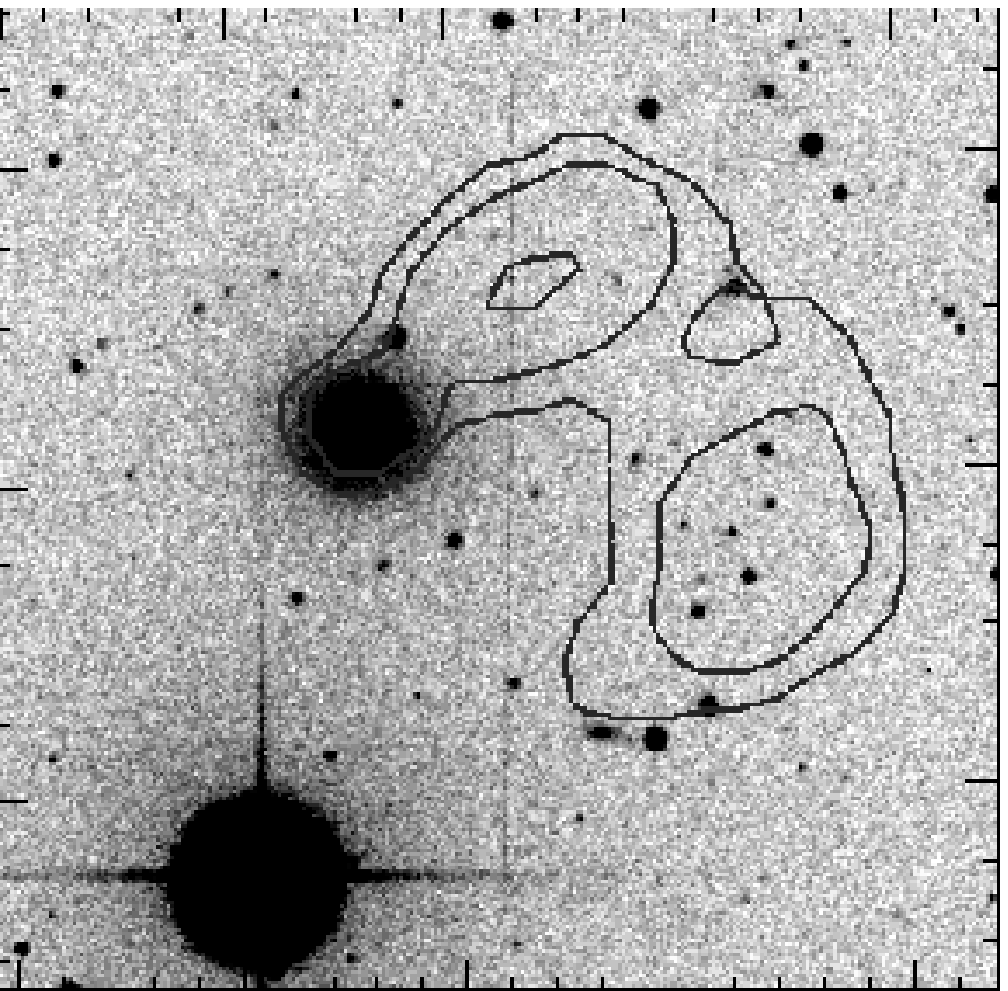}
\figcaption[NGC 2534]{\small
Contours of H{\sc i} emission are overlaid on the DSS image of 
NGC~2534.  This figure originally appeared in \citet{dave}.
\label{2534map}
}
\end{center}}

\vbox{
\begin{center}
\includegraphics[width=\textwidth]{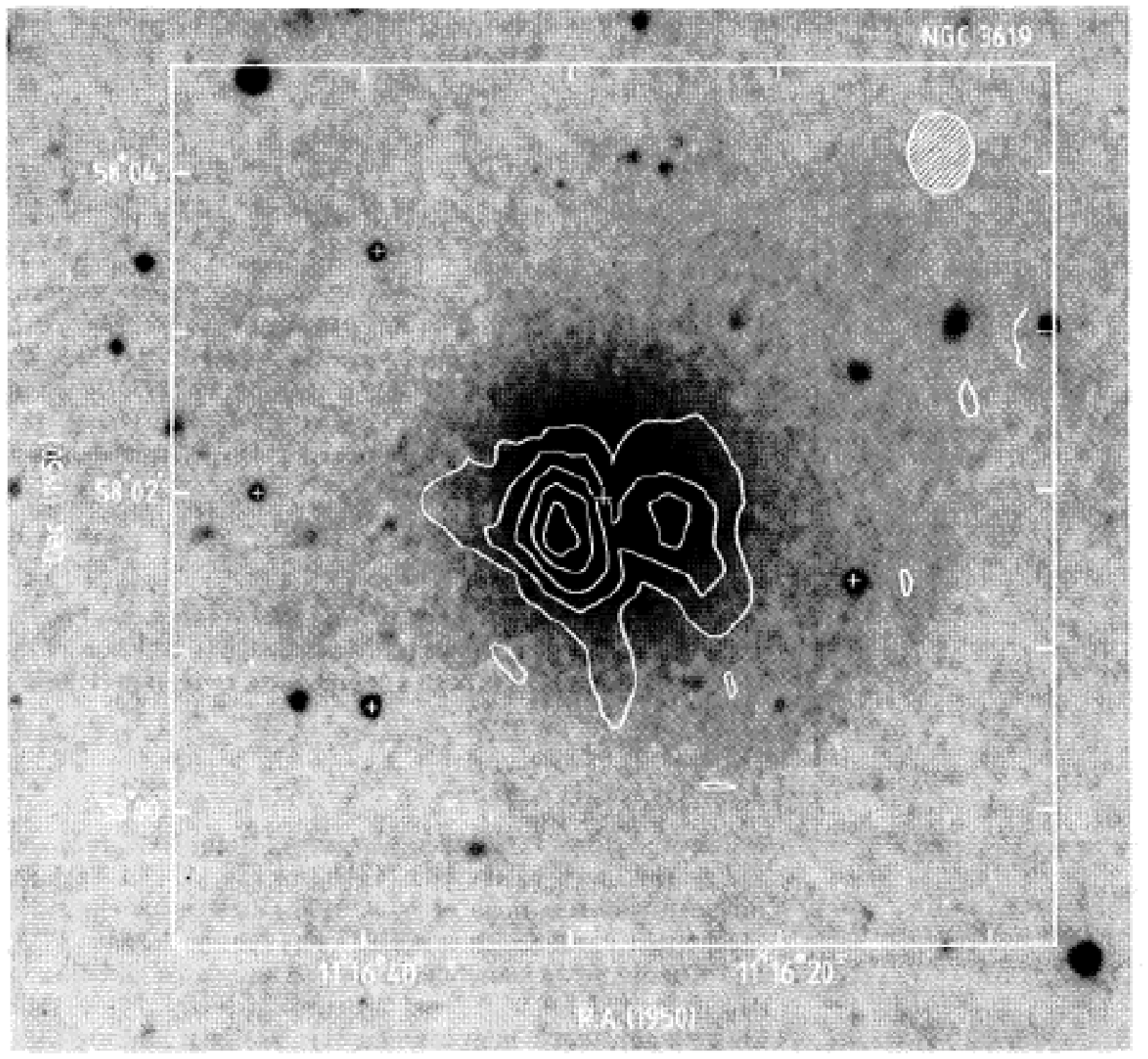}
\figcaption{\small
Contours of H{\sc i} emission are overlaid on the DSS image of 
NGC~3619.  This figure originally appeared in \citet{vdriel89}.
\label{3619map}
}
\end{center}}

\vbox{
\begin{center}
\includegraphics[0in,0in][8in,7in]{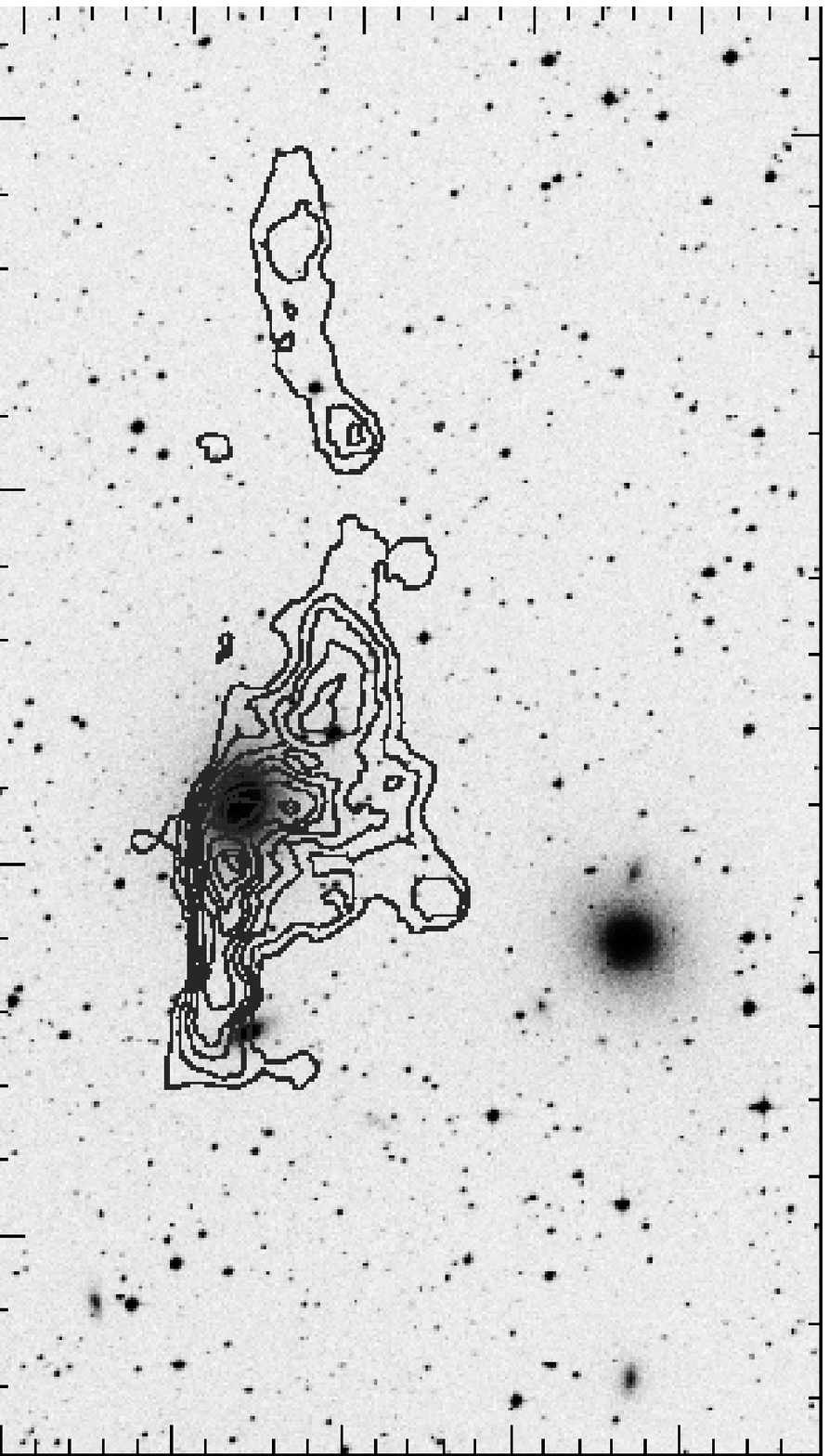}
\figcaption{\small
Contours of H{\sc i} emission are overlaid on the DSS image of 
NGC~5903.  This figure originally appeared in \citet{appleton}.
\label{5903map}
}
\end{center}}

\vbox{
\begin{center}
\includegraphics[width=\textwidth]{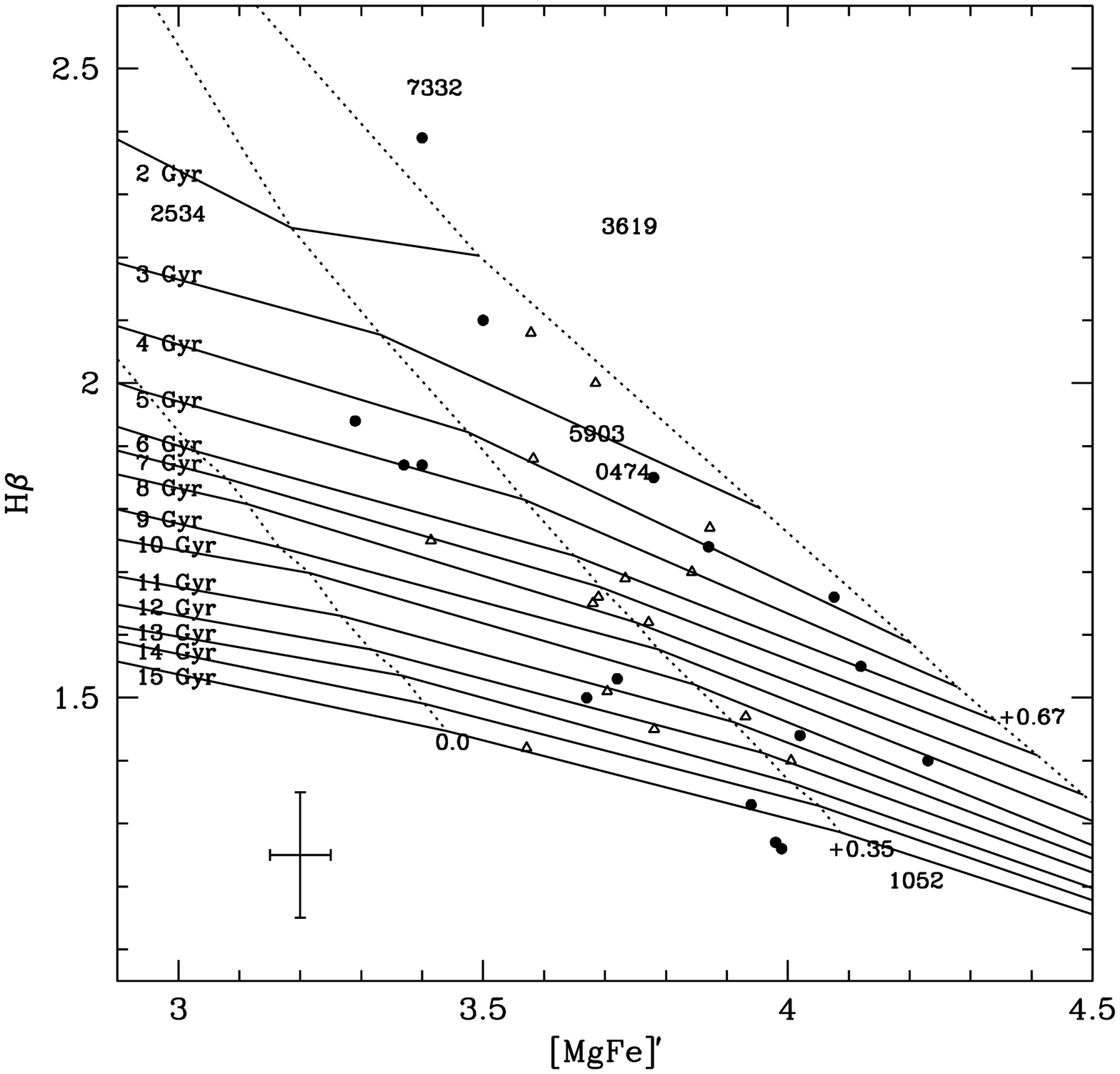}
\figcaption{\small
Galaxies from the merger remnant sample are plotted on the H$\beta$ vs. 
[MgFe]$^\prime$ plane (NGC numbers).  Also shown are galaxies from the
volume limited sample.  Solid circles are from Paper~I; open triangles are
from G93.  Models are from \citet{ch2tmb03}.  Typical error bars are in
the lower left.
\label{hbmr}
}
\end{center}}

\vbox{
\begin{center}
\includegraphics[width=\textwidth]{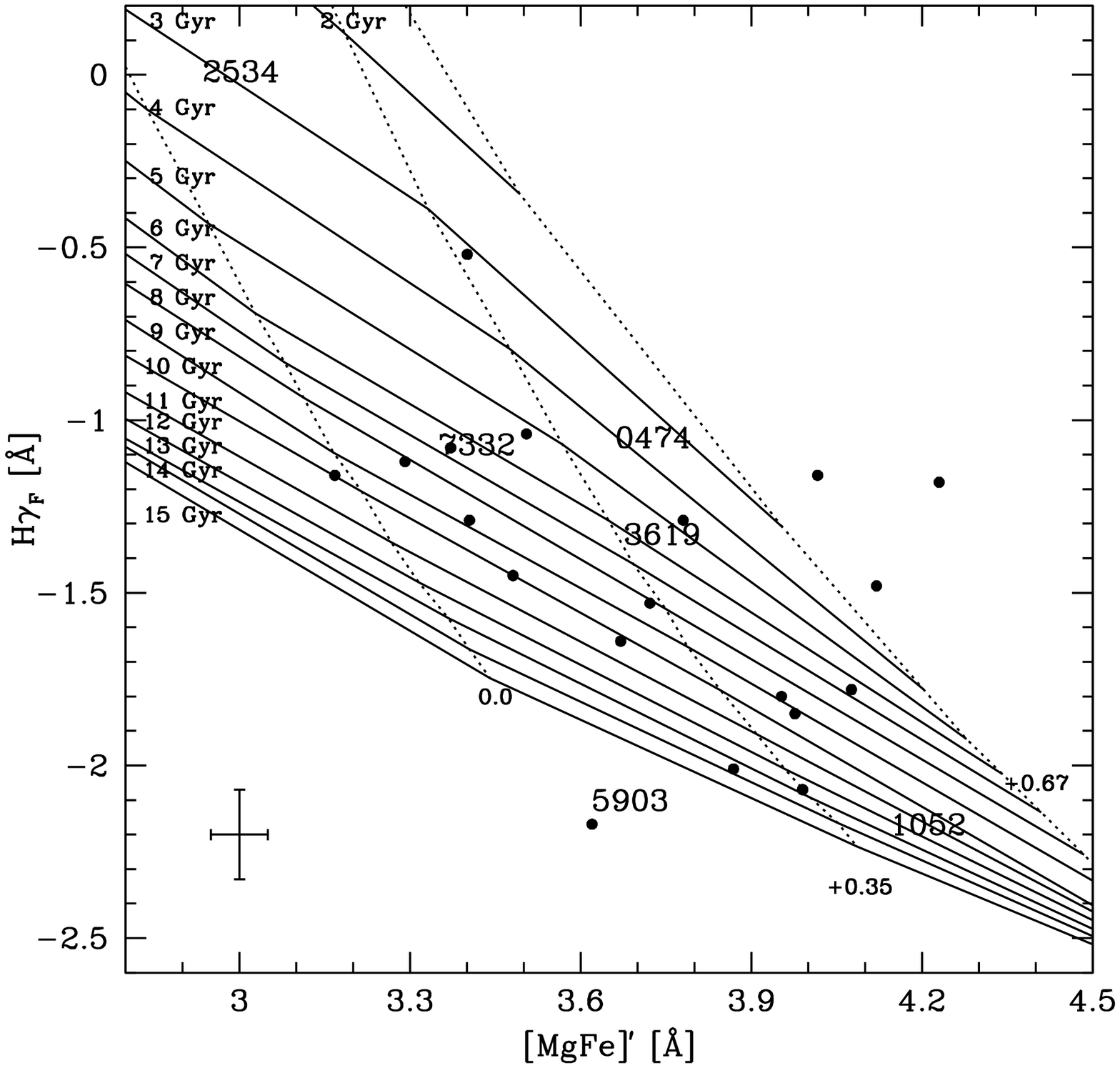}
\figcaption{\small
Galaxies from the merger remnant sample are plotted on the H$\gamma_{\rm F}$ 
vs. [MgFe]$^\prime$ plane (NGC numbers).  Also shown are galaxies from the
volume limited sample (black points; Paper~I).  Models are from 
\citet{tmk04}.  Typical error bars are in the lower left.
\label{hgmr}
}
\end{center}}

\vbox{
\begin{center}
\includegraphics[width=\textwidth]{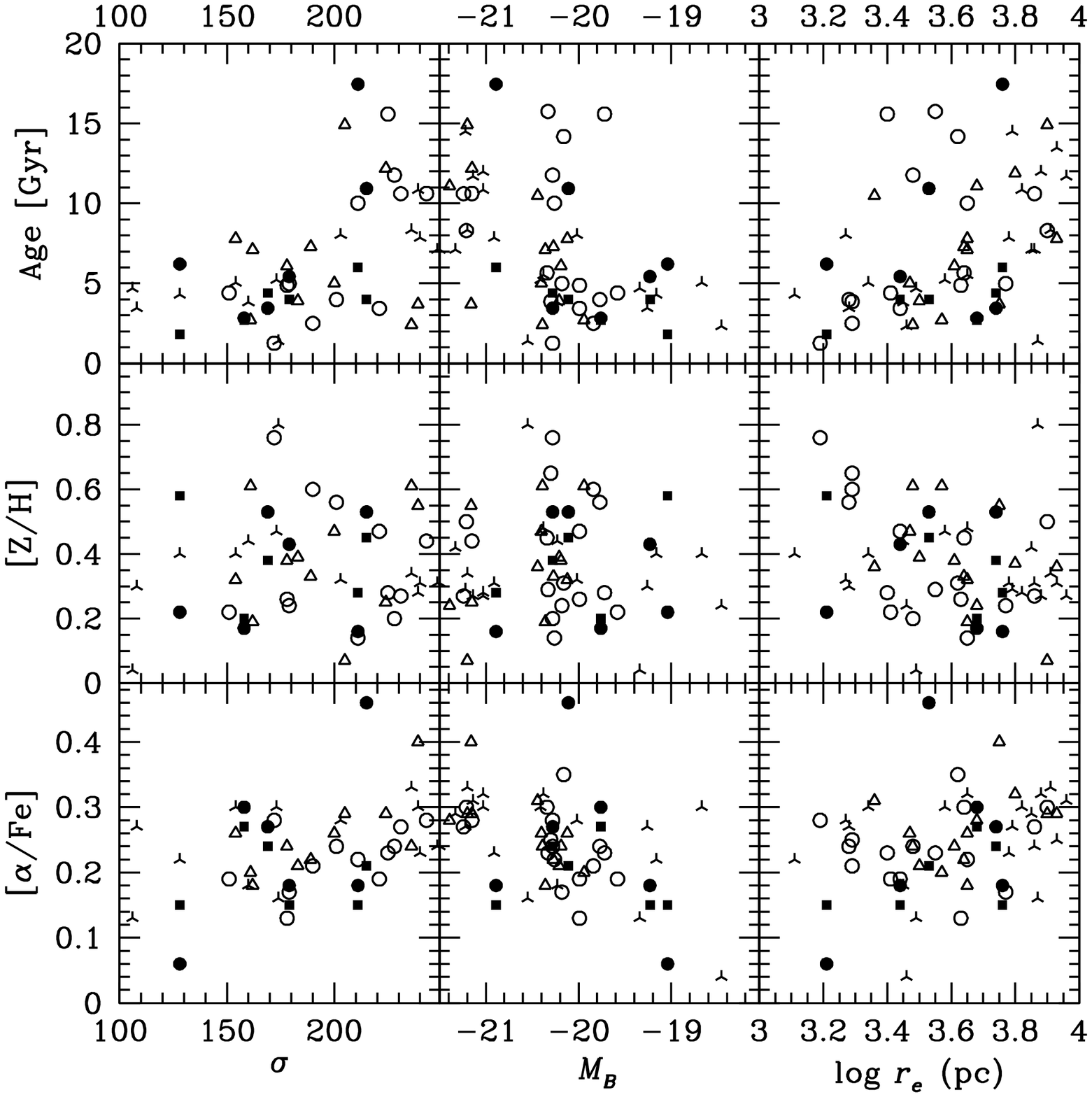}
\figcaption{\small
Comparison between parameters relating to the average stellar composition
(age, [Z/H], [$\alpha$/Fe]) and parameters relating to galaxy size ($\sigma$,
$M_B$, log$r_e$).  Open circles are galaxies from the volume-limited
sample (Paper~I), open triangles are G93 galaxies included in the
volume-limited sample, and three-pointed stars are the G93 galaxies not
included in the volume-limited sample.  The solid circles represent the merger
remnant sample as measured using the H$\gamma$ index SSP models, while the 
solid squares are the same galaxies as measured by the multi-index fitting 
procedure.  Contrary to expectations \citep{tgb99}, the 
merger remnant sample is distributed consistently with the 
volume-limited sample.
\label{asigmr}
}
\end{center}}

\vbox{
\begin{center}
\includegraphics[width=\textwidth]{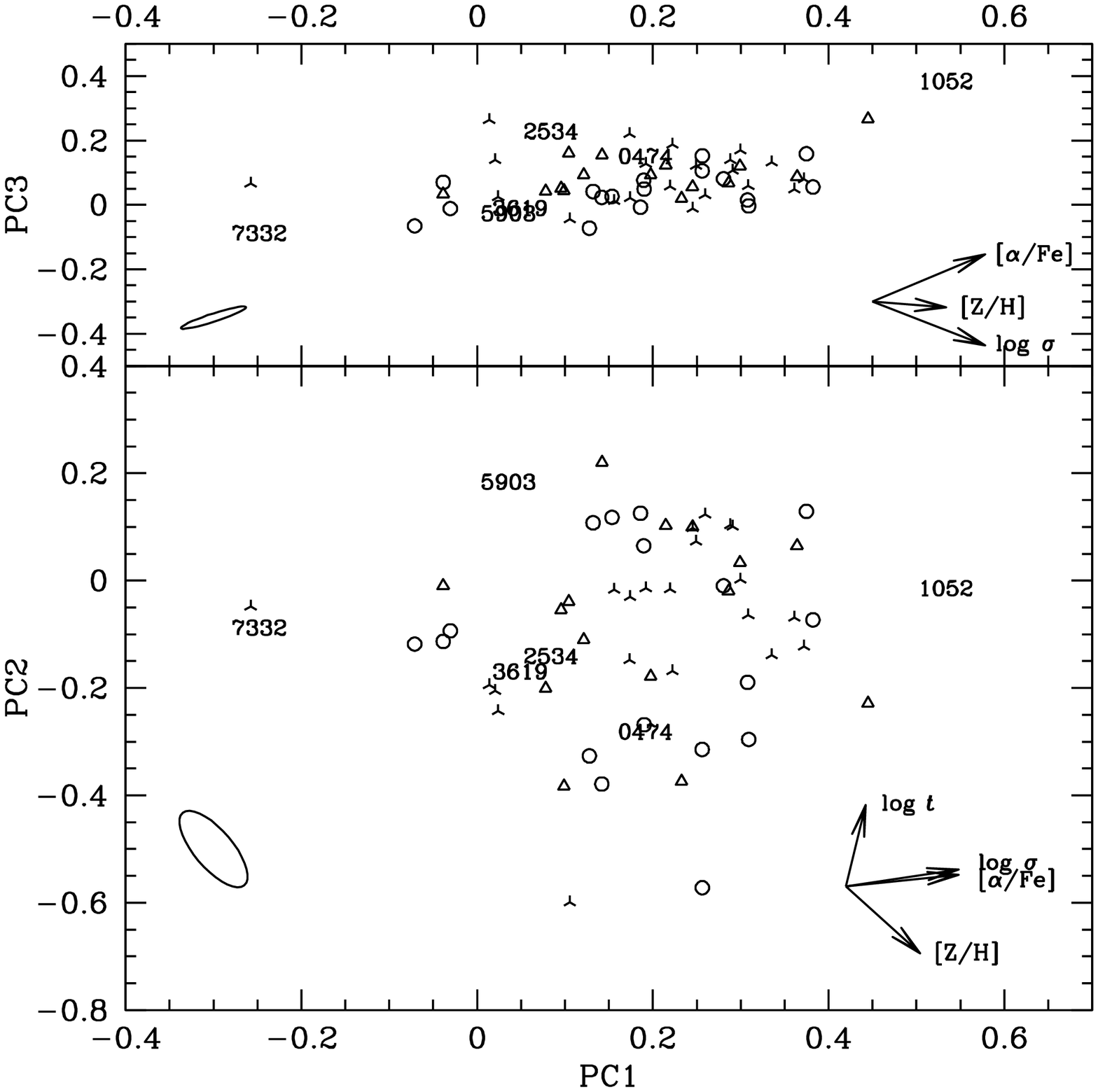}
\figcaption{\small
Merger remnant galaxies plotted on the metallicity hyperplane.  Open 
symbols and three-pointed stars 
are as in Fig.~\ref{asigmr}.  Merger remnant galaxies are plotted using
their NGC numbers.  Projections of each SSP parameter along principal
component axes are shown in the lower right.  PC1 depends primarily on 
velocity dispersion and $\alpha$-enhancement, PC2 depends primarily on 
age and metallicity, and PC3 measures deviations from the 
[$\alpha$/Fe]--$\sigma$ relation.  See Paper~I for details.  The merger 
remnant galaxies are generally consistent with the distribution of the 
volume-limited sample.  NGC~7332 has the low PC1 value expected 
of a merger remant \citep{tgb99}; however the galaxy is otherwise 
consistent with the volume-limited sample.
\label{pcmr}
}
\end{center}}

\vbox{
\begin{center}
\includegraphics[width=\textwidth]{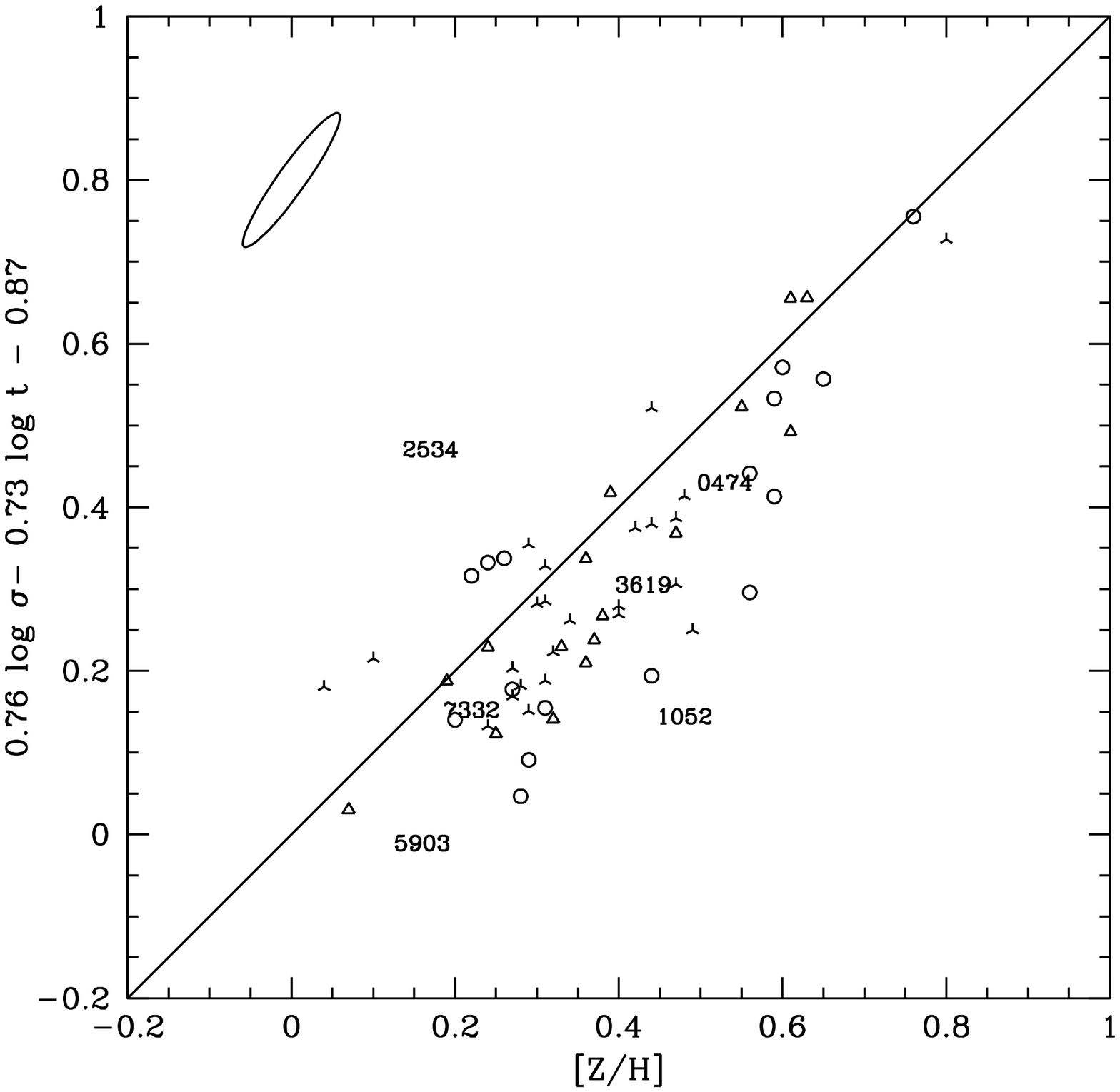}
\figcaption{\small
Merger remnant galaxies plotted on the $Z$-plane.  Points are as in 
Fig.~\ref{pcmr}.  The best-fit line from \citet{ch2tfwg2} is shown.  The
offset between this line and the locus of the volume-limited sample is
due to differences in models between this work and \citet{ch2tfwg2}; see
text for details.  The deviation of
NGC~2534 from the plane is plausibly explained by a recent starburst 
overlying an older stellar population that constitutes the majority of
the galaxy's stellar mass.
\label{zplanemr}
}
\end{center}}

\end{document}